\documentclass[sigconf]{acmart}
\AtBeginDocument{%
  }

\usepackage{multirow}
\usepackage{enumitem}

\copyrightyear{2026}
\acmYear{2026}
\setcopyright{cc}
\setcctype{by}
\acmConference[SIGIR '26]{Proceedings of the 49th International ACM SIGIR Conference on Research and Development in Information Retrieval}{July 20--24, 2026}{Melbourne, VIC, Australia}
\acmBooktitle{Proceedings of the 49th International ACM SIGIR Conference on Research and Development in Information Retrieval (SIGIR '26), July 20--24, 2026, Melbourne, VIC, Australia}
\acmDOI{10.1145/3805712.3808409}
\acmISBN{979-8-4007-2599-9/2026/07}

\begin{document}

\title{Synthetic Data Powers Product Retrieval for Long-tail Knowledge-Intensive Queries in E-commerce Search}



\author{Gui Ling}
\email{linggui.lg@alibaba-inc.com}
\affiliation{%
  \institution{Taobao \& Tmall Group of Alibaba}
  \city{Hangzhou}
  \country{China}
}

\author{Weiyuan Li}
\email{liweiyuan.lwy@alibaba-inc.com}
\affiliation{%
  \institution{Taobao \& Tmall Group of Alibaba}
  \city{Hangzhou}
  \country{China}
}

\author{Yue Jiang}
\email{jy270069@alibaba-inc.com}
\affiliation{%
  \institution{Taobao \& Tmall Group of Alibaba}
  \city{Hangzhou}
  \country{China}
}

\author{Wenjun Peng}
\email{pengwj@mail.ustc.edu.cn}
\affiliation{%
  \institution{Taobao \& Tmall Group of Alibaba}
  \city{Hangzhou}
  \country{China}
}

\author{Xingxian Liu}
\email{liuxingxian@bupt.edu.cn}
\affiliation{%
  \institution{Taobao \& Tmall Group of Alibaba}
  \city{Hangzhou}
  \country{China}
}

\author{Dongshuai Li}
\email{lidongshuai.lds@alibaba-inc.com}
\affiliation{%
  \institution{Taobao \& Tmall Group of Alibaba}
  \city{Hangzhou}
  \country{China}
}

\author{Fuyu Lv}
\authornote{Corresponding author.}
\email{fuyu.lfy@alibaba-inc.com}
\affiliation{%
  \institution{Taobao \& Tmall Group of Alibaba}
  \city{Hangzhou}
  \country{China}
}

\author{Dan Ou}
\email{oudan.od@alibaba-inc.com}
\affiliation{%
  \institution{Taobao \& Tmall Group of Alibaba}
  \city{Hangzhou}
  \country{China}
}

\author{Haihong Tang}
\email{piaoxue@taobao.com}
\affiliation{%
  \institution{Taobao \& Tmall Group of Alibaba}
  \city{Hangzhou}
  \country{China}
}

\renewcommand{\shortauthors}{Gui Ling et al.}

\begin{abstract}
Product retrieval is the backbone of e-commerce search: for each user query, it identifies a high-recall candidate set from 
billions of items, laying the foundation for high-quality ranking and user experience.
Despite extensive optimization for mainstream queries, existing systems still struggle with long-tail queries, especially knowledge-intensive ones. These queries exhibit diverse linguistic patterns, often lack explicit purchase intent, and require domain-specific knowledge reasoning for accurate interpretation. They also 
suffer from a shortage of reliable behavioral logs, which makes such queries a persistent challenge for retrieval optimization.

To address these issues, we propose an \emph{efficient data synthesis framework} tailored to retrieval involving long-tail, knowledge-intensive queries. The key idea is to implicitly distill the capabilities of a powerful offline query-rewriting model into an efficient online retrieval system.
Leveraging the strong language understanding of LLMs, we train a multi-candidate query rewriting model with multiple reward signals and capture its rewriting capability in well-curated query–product pairs through a powerful offline retrieval pipeline. This design mitigates distributional shift in rewritten queries, which might otherwise limit incremental recall or introduce irrelevant products.
Experiments demonstrate that without any additional tricks, simply incorporating this synthetic data into retrieval model training leads to significant improvements. Online Side-By-Side (SBS) human evaluation results indicate a notable enhancement in user search experience.
\end{abstract}

\maketitle

\begin{figure*}[t]
    \centering
    \includegraphics[width=0.95\linewidth]{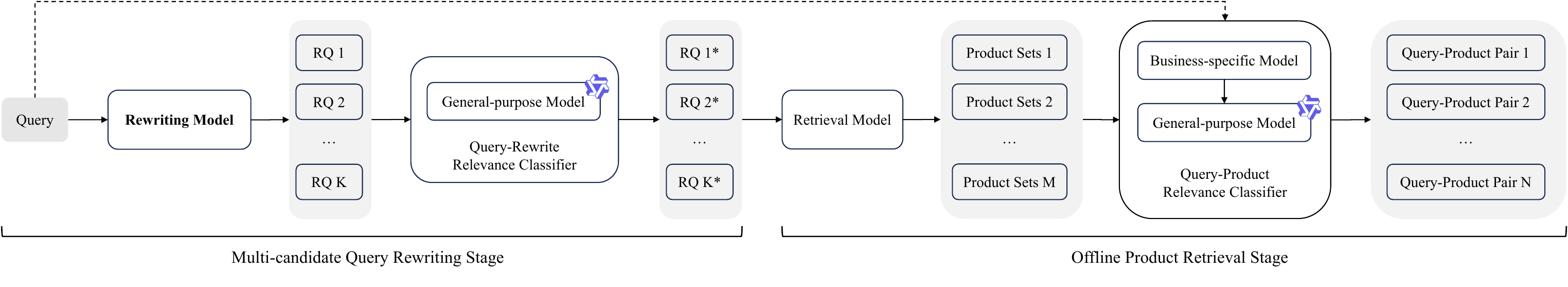}
    \caption{Our proposed data synthesis framework targeting long-tail knowledge-intensive queries. The framework comprises two stages: (1) Multi-Candidate Query Rewriting and (2) Offline Product Retrieval. 
    Query rewriting is the core component of the framework. The retrieval model and the business-specific query-product relevance model are optimized production models aligned with online serving system, while the query–rewrite relevance model and the general-purpose query-product relevance model are open-source LLMs without domain-specific fine-tuning.
    }
    \label{fig:pipeline} 
\end{figure*}

\section{Introduction}
Large-scale e-commerce product retrieval systems such as Taobao and Amazon are essential for helping users efficiently discover relevant items from vast and dynamic inventories. Over the years, retrieval models have evolved from traditional inverted index and BM25-based methods~\cite{svore2009machine} to more sophisticated semantic matching learning approaches~\cite{huang2013learning, yi2019sampling, reimers2019sentence}. Recently, the integration of large language models (LLM) has significantly advanced the field, such as LLM-powered dense retrieval~\cite{zhang2025qwen3, su2022one} and LLM-based generative retrieval~\cite{li2024generative}. 
These approaches leverage the rich world knowledge and strong generalization capabilities of LLMs, offering promising improvements in recall performance across diverse query types.

Despite these advancements, 
we find that retrieval quality remains unsatisfactory for long-tail, knowledge-intensive queries. These queries often express complex user needs involving specific functional requirements or product descriptions, or require domain-specific knowledge or multi-hop reasoning capabilities, such as "What are the ingredients of 'Ants Climbing a Tree' (a Chinese dish)?". 
Their low frequency in user logs results in little labeled training data, and
manual annotation is both time-consuming and costly. As a result, even state-of-the-art retrieval models struggle to understand and match such queries to the correct products. 
A common way to improve long-tail retrieval is query rewriting~\cite{farzana2023knowledge, qiu2021query, nguyen2025rl, peng2024large, feng2025complicated}, which maps rare queries to more standard, frequent forms that existing models can better handle.
However, relying solely on rewriting is insufficient to address the core challenge. 
The fundamental limitation lies not only in query understanding but also in \emph{the lack of corresponding query-product pairs} needed to train the retrieval model effectively. Even if a rewriting model produces a semantically equivalent
query, the downstream retriever may still fail to return the correct item if it has never been exposed to similar product associations during training. 
Moreover, rewrites can inadvertently omit critical constraints or introduce intent biases, especially when the original query involves subtle specifications or rare terminology. Without high-quality labeled data that connects rewritten queries to ground-truth items, improvements in query rewriting do not translate into better retrieval results.

Motivated by this observation, we propose an efficient data synthesis framework that distills offline query-rewriting capability into an efficient online retrieval system, by generating high-quality and semantically faithful training examples for retrieval tasks involving long-tail, knowledge-intensive queries.
Our approach centers on optimizing a strong offline multi-candidate query rewriting model, 
which is expected to (i) preserve semantic consistency to avoid meaning drift, and (ii) align with the retriever’s data distribution to improve efficiency. To this end, we design a multi-reward optimization framework: Query Semantic Relevance (QSR), Product-side Distribution Alignment (PDA), and Diversity across multiple candidates.
Even with these optimizations, unavoidable failure cases remain, making it difficult to deploy the rewriting model directly for recall items expansion in production. We therefore replicate the online retrieval stack to build an offline retrieval pipeline composed of stronger LLM-enhanced models.
In this pipeline, the retrieval module uses rewritten queries to retrieve products and generate large-scale query–product pairs, 
while the relevance module scores each pair and retains 
those deemed highly relevant.
Online Side-By-Side (SBS) human evaluations show that data augment with the synthetic data improves the Query Goodrate by up to 8.62pt.

\section{Methods}

The architecture of our data synthesis framework is illustrated in Figure \ref{fig:pipeline}, which consists of a Multi-candidate Query Rewriting stage and an Offline Product Retrieval stage. 

\begin{figure}[t]
    \centering
    \includegraphics[width=1\linewidth]{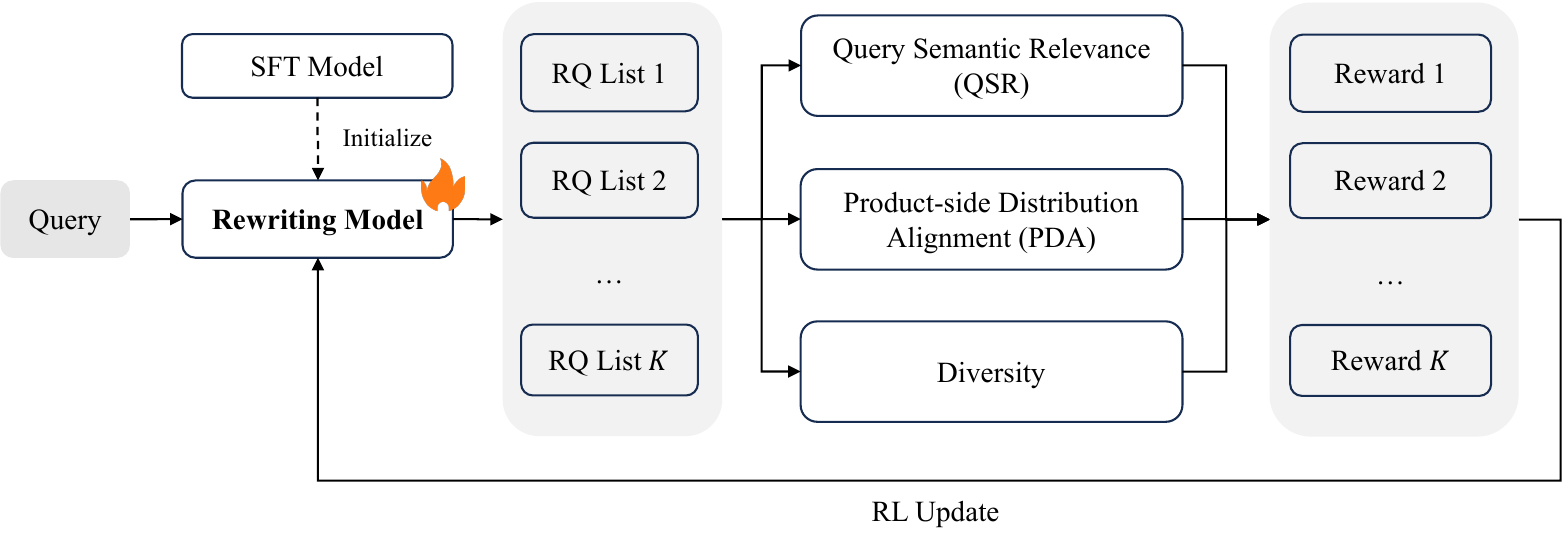}
    \caption{A multi-reward design for rewriting model optimization. In each iteration, multiple responses (each containing a rewrite list) are sampled and scored on relevance, alignment, and diversity. The three rewards are then weighted and summed to form the final reward for model updates.}
    \label{fig:rewrite} 
\end{figure}

\noindent\textbf{Rewriting Model Optimization}. 
In general e-commerce scenarios,
the rewriting alternative expressions are used together with the original query in a candidate expansion phase to improve recall coverage.
Rewriting behavior is designed to be conservative to avoid generating irrelevant variants that could lead to excessive noise in downstream.
For long-tail, knowledge-intensive queries, such as "What are the ingredients of ’Ants Climbing a Tree’?", simple expansion is insufficient because retrieving relevant items requires reasoning or external knowledge. 
In such cases, the purpose of rewriting shifts from \textit{expanding} the original query to \textit{replacing} it with a set of candidate rewrites, each capturing a plausible sub-intent or interpretation. We refer to this approach as \emph{multi-candidate rewriting}. The goal is to generate a diverse yet faithful set of rewrites that collectively cover the full intent space of the original query.

The model follows a two-stage training paradigm: Supervised Fine-Tuning (SFT) followed by Reinforcement Learning (RL).
We begin by using open-source Qwen3-30B-A3B~\cite{yang2025qwen3} to generate a small set of rewrite pairs from real user queries. These synthetic samples are manually reviewed and cleaned according to business rules. This curated dataset is then used to train the baseline rewriting model via SFT, giving it fundamental capabilities in producing semantically consistent rewrites.
We introduce a multi-reward scheme during RL training as illustrated in Figure \ref{fig:rewrite}. 
The reward function consists of three complementary components:
\begin{itemize} 
\item \textbf{Query Semantic Relevance (QSR)}. This measures how well the rewritten query preserves the intent of the original. To avoid reliance on costly human annotation, we reuse the human-labeled \textit{<query, rewrite>} pairs (positive vs. negative) from the rewriting model’s SFT data to train an LLM-based binary classifier. During RL training, the confidence derived from the model’s output logits $z$ is used as the reward signal as Equation \ref{eq:score1},
\begin{equation}
R_{\text{QSR}} = p_{\text{yes}} = \frac{\exp(z_{\text{yes}})}{\exp(z_{\text{yes}}) + \exp(z_{\text{no}})}
\label{eq:score1}
\end{equation}

\item \textbf{Product-side Distribution Alignment (PDA)}. To encourage rewrites to align with actual product-side language (e.g., titles, tags), 
We perform continued pre-training (CPT) on a LLM using product-domain corpora to obtain a model of product expression distribution. For each rewritten query, we compute its perplexity (PPL) under this model; lower PPL indicates better alignment with real product language.
In RL, we compute the alignment score using PPL as Equation \ref{eq:score2},
\begin{equation}
R_{\text{PDA}} = \frac{1}{\text{PPL}} = \exp\left(\frac{1}{T} \sum_{t=1}^{T} \log p(x_t \mid x_{<t})\right)
\label{eq:score2}
\end{equation}

\item \textbf{Diversity Reward}. Since we generate multiple rewrites per query, we promote diversity among them to capture different aspects of the original intent. We compute pairwise dissimilarity across the generated rewrites using n-gram overlap and derive a diversity score as Equation \ref{eq:score3},
\begin{equation}
R_{\text{Diversity}} = \frac{1}{\binom{k}{2}} \sum_{i < j} \left(1 - \frac{|\mathbf{n}_i \cap \mathbf{n}_j|}{|\mathbf{n}_i \cup \mathbf{n}_j|}\right)
\label{eq:score3}
\end{equation}
where $ \mathbf{n}_i $ denotes the set of character $ n $-grams in rewrite $ r_i $ with $ n=2 $, and $ k = 5 $ denotes the number of rewrites considered.
\end{itemize}

Note that we employ a standard format reward commonly used in RL, and the multi-reward signal is activated only when the format is correct.
Considering that all the three rewards are continuous rather than discrete, we optimize the model using the REINFORCE++ algorithm~\cite{hu2025reinforce++}, a policy gradient method in reinforcement learning. The overall training objective is defined as Equation \ref{eq:objective},
\begin{equation}
\mathcal{L}_{\text{RL}} = -\mathbb{E}[\, R_{\text{total}} \cdot \log p(\text{rewrites}|\text{query}) \,]
\label{eq:objective}
\end{equation}
where 
\begin{equation}
R_{\text{total}} = \alpha R_{\text{QSR}} + \beta R_{\text{PDA}} + \gamma R_{\text{Diversity}}
\end{equation}
and $\alpha$, $\beta$ and $\gamma$ are hyper-parameters, heuristically set to 1.0, 0.5, and 0.1, respectively. 

\noindent\textbf{Query-Rewrite Relevance Classifier}. 
In the subsequent Offline Product Retrieval stage, we observe a pronounced data explosion. To reduce the inference burden on the relevance module, we introduce a rewriting classifier after the rewriting model to filter out semantically irrelevant rewrites. This classifier is initialized from a general-purpose LLM and performs conservative, high-recall filtering. We design a lightweight prompt based on a summary of business rules and ask the model to output one of three labels: \emph{relevant}, \emph{partially relevant}, or \emph{irrelevant}. Any rewrite labeled as \emph{irrelevant} is removed.

\noindent\textbf{Retrieval Model}. 
We use the same dense retriever as deployed online, an in-house, optimized 3B LLM-based retrieval model trained on Taobao historical user behavior data~\cite{liu2025taosearchemb}. In implementation, each rewritten query is encoded into a dense embedding by the retriever and matched against the item corpus via Approximate Nearest Neighbor (ANN) to retrieve the top-200 most similar items.

\noindent\textbf{Query-Product Relevance Classifier}. 
The relevance module is critical to the quality of the synthesized data. We employ an in-house, optimized 42B MoE-based relevance model, trained on large-scale query–product pairs annotated by business rules, which exhibits strong generalization on long-tail, knowledge-intensive queries~\cite{yang2025taosr}. In practice, each item retrieved by a rewritten query is paired with the original query to form a \textit{<query, product>} instance, which is then fed into the relevance module to produce a relevance score, where higher scores indicate stronger relevance.

Relying solely on a business-specific relevance model is insufficient. Since it is fitted to business-rule annotations and it can produce false positives that are easily distinguishable under general semantic reasoning. 
We find that it may overemphasize lexical overlap and incorrectly mark pairs as relevant based on term matching.
For instance, the query “portable power station for camping” and a product titled “power station maintenance tool” share surface keywords but differ substantially in user intent.
To address this issue, we introduce a general semantic relevance model, typically a strong open-source LLM, to perform a semantic assessment and catch false positives induced by business rules. To avoid discarding potentially valid pairs, we prompt the general model with a conservative, high-recall filtering, as in the rewriting classifier.

\begin{table*}[t]
    \centering
    \caption{The Impact of Synthetic Data Augmentation on Retrieval Performance}
    \begin{tabular}{llllllll}
    \toprule
        \multirow{2}{*}{Query Type \footnotesize{(Case)}} & \multicolumn{2}{l}{Item Goodrate} & \multicolumn{2}{l}{Query Goodrate@10} & \multicolumn{2}{l}{Query Goodrate@100} \\
        ~ & Baseline & w/ Synth. & Baseline & w/ Synth. & Baseline & w/ Synth. \\ \hline
        Q\&A\footnotesize{ (What medicine can turn hair black?)} & 81.85\% & 87.19\% \footnotesize{(+5.34pt)} & 96.31\% & 97.74\% \footnotesize{(+1.43pt)} & 87.73\% & 91.36\% \footnotesize{(+3.63pt)} \\ 
        Alternative\footnotesize{ (Affordable alternatives to Converse shoes)} & 23.45\% & 30.08\% \footnotesize{(+6.63pt)} & 53.43\% & 60.73\% \footnotesize{(+7.30pt)} & 21.57\% & 28.00\% \footnotesize{(+6.43pt)} \\ 
        Negative\footnotesize{ (Short sleeve shirt that doesn’t attract hair)} & 48.58\% & 61.53\% \footnotesize{(+12.95pt)} & 83.43\% & 88.88\% \footnotesize{(+5.54pt)} & 51.01\% & 64.16\% \footnotesize{(+13.15pt)} \\ 
        Knowledge\footnotesize{ (Best GPU for Black Myth: Wukong)} & 60.58\% & 71.31\% \footnotesize{(+10.73pt)} & 84.61\% & 89.74\% \footnotesize{(+5.13pt)} & 64.60\% & 74.20\% \footnotesize{(+9.60pt)} \\ 
        \bottomrule
    \end{tabular}
    \label{table:mainresult}
\end{table*}

\section{Experiments}

\subsection{Experiment Setup}

\textbf{Data Synthesis Pipeline Overview}. Our full data synthesis pipeline integrates multiple specialized models: The \textit{query rewriting model} is initialized from Qwen3-30B-A3B~\cite{yang2025qwen3} and fine-tuned on domain-specific rewriting tasks. 
During training, we first fine-tune the model on 5K manually annotated rewriting examples, and then apply multi-reward RL to further optimize it until the rewards largely converge.
During inference, the model generates a list of rewrites for each sample, containing at least 5 candidate rewritten queries.
The \textit{dense retrieval model} is derived from and optimized based on our in-house Tbstars-3B architecture. The \textit{business-specific relevance model} is a high-capacity model built on Tbstars-42B-A3B. Both the \textit{rewrite classifier} and the \textit{general-purpose relevance classifier} directly use the open-source Qwen3-30B-A3B model without additional fine-tuning, serving as strong off-the-shelf filters. We collected approximately 20M pairs from 400K queries.

\noindent\textbf{Evaluation Dataset and Metrics}. 
To facilitate a detailed evaluation of model behavior, we classify the long-tail, knowledge-intensive queries into four types based on their semantic and reasoning characteristics: 
\begin{itemize}
\item \textbf{Question-Answering}: Queries phrased as questions requiring factual or explanatory responses. 
\item \textbf{Affordable Alternative}: Queries seeking affordable alternatives to premium products. 
\item\textbf{Negative Expressions}: Queries that include negative constraints. 
\item\textbf{General Knowledge Type}: All other knowledge-intensive queries that do not fall into the above three types but still require external knowledge.
\end{itemize}

For offline evaluation, we sample 500 queries per query type from Taobao’s online search logs to reflect the real-world query distribution. For each query, we retrieve the top-200 products using the retrieval model and manually annotate query–item relevance.
We evaluate retrieval quality using \textit{Item Goodrate} and \textit{Query Goodrate@N}. \textit{Item Goodrate} measures item-level performance by averaging, across queries, the proportion of highly relevant items in the retrieved set. \textit{Query Goodrate@N} is defined as the fraction of queries for which at least \(N\) relevant items are retrieved within the top-200 results, capturing per-query coverage of relevant content.

In online A/B testing, we similarly sample 500 queries per query type and evaluate the relevance of the top-10 results on the Taobao Search Results Page (SRP). We conduct a Side-By-Side (SBS) human evaluation: SRP results from the control and treatment buckets are displayed side by side, and annotators perform blinded evaluations.
We evaluate the SRP results using three key metrics:
\textit{Item Goodrate}, \textit{Query Goodrate} and \textit{GSB (Good/Same/Bad)}. Query Goodrate is a page-level relevance metric measuring the proportion of queries whose SRP results are rated "Good" or "Mid" based on the relevance of displayed items. 
GSB measures the relative quality of SRP results, with \(\text{GSB}+x\%\) indicating that \(x\%\) of the test results are preferred over the baseline, capturing the relative improvement.

\subsection{Offline Evaluation}

Table~\ref{table:mainresult} reports the retrieval gains from synthetic data augmentation; both settings are trained on the same scale of 200M data using Tbstars-3B. The baseline does not include any synthetic data tailored to long-tail queries, while the augmented setting incorporates such data. The data augmented version consistently improves \textit{Item Goodrate} and \textit{Query Goodrate@N} across all query types. The largest \textit{Item Goodrate} gains come from Negative and Knowledge queries (+12.95pt and +10.73pt). Negative queries are especially difficult for standard semantic matching; for example, “non-wooden wardrobe” often still retrieves wooden wardrobes.
For Knowledge queries, a substantial portion concerns outfit coordination. For example, given the query “what coat pairs with a blue dress?”, the baseline model frequently retrieves “blue coats,” which are only partially relevant and reflect a shallow interpretation of the underlying intent.

\noindent\textbf{Evaluation of Synthetic Data}. 
We randomly sample 1,000 query-product pairs from each query type and conduct human annotation for evaluation. 
As shown in Table~\ref{table:data}, the accuracy exceeds 80\% for three of the four types, and the Q\&A type achieves an accuracy above 90\%. 
The only exception is the Alternative, which falls below 80\% (64.5\%), consistent with our earlier results. This is largely because it requires strong domain knowledge and timely information, which the model’s inherent knowledge cannot reliably provide.

\begin{table}[t]
    \centering
    \caption{Accuracy Assessment of Synthesized Data}
    \begin{tabular}{ccccc}
    \toprule
        ~ & Q\&A & Alternative & Negative & Knowledge \\ \hline
        Accuracy & 90.6\% & 64.5\% & 81.6\% & 86.0\% \\ \bottomrule
    \end{tabular}
    \label{table:data}
\end{table}

\begin{table}[t]
    \centering
    \caption{Evaluation of Query Rewriting Optimization}
    \begin{tabular}{cccc}
    \toprule
        ~ & Accuracy & PDA PPL & Item Goodrate \\ \hline
        Baseline & 39.6\% & - & - \\ 
        Opt. w/o PDA & 67.5\% & 1407.4 & 54.1\% \\ 
        Opt. w/ PDA & 70.9\% & 134.5 & 72.8\% \\ \bottomrule
    \end{tabular}
    \label{table:rewrite}
\end{table}

\begin{table}[t]
    \centering
    \caption{Online SBS Human Evaluations}
    \begin{tabular}{cccc}
    \toprule
        Query Type & Item Goodrate & Query Goodrate & GSB \\ \hline
        Q\&A &  +5.12pt &  +7.45pt &  +19.71\% \\ 
        Alternative &  +5.00pt &  +8.62pt &  +27.66\% \\ 
        Negative &  +6.97pt &  +7.65pt &  +31.59\% \\ 
        Knowledge &  +2.58pt &  +5.82pt &  +15.66\% \\ \bottomrule
    \end{tabular}
    \label{online_result}
\end{table}
\noindent\textbf{Evaluation of Rewriting Model}. 
The rewriting model’s performance strongly affects data synthesis efficiency. To quantify this effect, we compare the off-the-shelf Qwen3-30B-A3B (baseline) with our task-optimized variant. We conduct human evaluation on rewrites from both models and measure semantic consistency by accuracy. As shown in Table~\ref{table:rewrite}, the optimized model substantially improves rewriting accuracy.
Rewriting quality is influenced by multiple factors; here we focus on the impact of our proposed PDA. Incorporating PDA markedly reduces perplexity and substantially improves the Goodrate of retrieved products, indicating better alignment with product-side language patterns.

\subsection{Online Evaluation}

We developed an LLM-based online retrieval pipeline that uses a query understanding model to route knowledge-intensive queries and improves the search experience with LLM-based retrieval and relevance models.
We conducted online A/B testing on the Taobao platform for over four weeks, allocating 10\% of long-tail-query traffic. Our optimization objective is query–product relevance , and we improved user experience without hurting order count, with user experience measured via SBS human evaluation.
As shown in Table~\ref{online_result}, the experimental group, augmented with synthetically generated data,
achieves significant improvements across all three evaluation metrics and all four query types. The largest gain in \emph{Item Goodrate} is observed on the Negative query type (+6.97 pts), while the highest improvement in \emph{Query Goodrate} occurs on the Alternative type (+8.62 pts). These query categories often follow consistent expression patterns and represent fine-grained user intents, which are challenging for traditional retrieval systems to handle due to their semantic complexity. 

\section{Conclusion}

We propose an efficient data synthesis framework to mitigate training data scarcity for product retrieval involving long-tail, knowledge-intensive queries. By training the rewriting model with a multi-reward framework and capturing the rewriting model’s capability as high-quality query–product pairs via an offline retrieval pipeline, we substantially improves the retrieval experience for such queries while avoiding the latency cost and rewrite instability associated with direct deployment of the rewriting model.

\noindent\textbf{Limitations}. Our framework is limited to textual knowledge and cannot effectively handle queries that depend on multimodal signals, such as requests for “same-style clothing”. In addition, although our method improves retrieval performance on “alternative” queries, the absolute performance remains suboptimal. For such queries, relying solely on the rewriting model’s parametric knowledge is insufficient.


\bibliographystyle{ACM-Reference-Format}
\balance
\bibliography{sigir26}










\end{document}